# Platinum Ditelluride based Field Effect Transistors for highly sensitive H$_2$S sensing

Areej Shahid
*Department of Electrical and Electronic Engineering, The University of Melbourne, Parkville, Victoria, Australia*
Email: areejs@student.unimelb.edu.au

Ranjith R Unnithan
*Department of Electrical and Electronic Engineering, The University of Melbourne, Parkville, Victoria, Australia*
Email: r.ranjith@unimelb.edu.au

Jackson Gum
*Department of Materials Science and Engineering, Monash University*
Clayton, Victoria 3800, Australia;
Email: Jackson.Gum@monash.edu

Sudha Mokkapati
*Department of Materials Science and Engineering, Monash University*
Clayton, Victoria 3800, Australia;
Email: sudha.mokkapati@monash.edu

*Abstract–Two-dimensional materials, specifically transition metal dichalcogenides, for highly sensitive gas sensing are emerging as effective detection technology. Platinum ditelluride (PtTe$_2$) is an intriguing material in the realm of field-effect transistors (FETs) due to its unique electronic properties. In this study, the CVD-synthesized PtTe$_2$ was functionalized with Au-Pd thin film for analyte-specific (H$_2$S) sensing for better toxic gas sensitivity and selectivity. It was concluded that PtTe$_2$, in combination with appropriate metal or oxide decorations, had great potential for ultrasensitive and selective, real-time gas sensor applications.*
Keywords—gas sensor, FET, H$_2$S, metal decoration, platinum ditelluride

## 1. Introduction

In recent decades, the detection and monitoring of toxic gases, such as volatile organic compounds (VOCs), hydrogen sulfide (H$_2$S), ammonia (NH$_3$), and nitrogen dioxide (NO$_2$) have become paramount in ensuring environmental/industrial safety and public health. Specifically, H$_2$S poses significant industrial, household, and laboratory threats [1]. Commercially available gas sensors work on chemiresistive gas sensing mechanism with on-chip heaters. They are, therefore, bulky, consume much power, are cross-sensitive, and have calibration inconsistencies. [2]

Field-effect transistors (FETs) have emerged as promising candidates for gas sensing applications due to their sensitivity, rapid response, and ease of integration into electronic systems [3,4]. Two-dimensional materials for highly sensitive gas sensing are emerging as effective detection technology [5]. Platinum ditelluride (PtTe$_2$) is a promising material for making channels in FETs due to its unique electronic properties such as high carrier mobility. It is classified as a Type-II Dirac semimetal, characterized by its complex band structure, where conduction and valence bands touch at isolated points in momentum space [6]. This topological property is particularly advantageous for FETs. In PtTe$_2$-based FETs, the gate voltage applied to the material allows precise control over carrier density. This means the transistor can be turned on or off by tuning the gate voltage, making it suitable for digital applications. PtTe$_2$ offers high carrier mobility owing to the large surface-to-volume ratio, a critical factor for FETs, enabling fast switching and excellent device performance. PtTe$_2$ FETs have the potential to revolutionize electronics by delivering high-speed and high-frequency capabilities, especially by varying the thickness [7]. Researchers continue to investigate PtTe$_2$ and similar materials, making it an exciting field of study with the promise of groundbreaking sensing applications in the semiconductor industry.

This paper utilizes the underlying mechanisms of FET-based gas sensing, recent technological developments in PtTe$_2$, and surface functionalization using metal decorations (Au and Pd) for targeted H$_2$S sensing. We demonstrate the two-dimensional (2D) material synthesis with a unique metallic functionalization, followed by studying its morphological, electrical, and gas-sensing characteristics. The selectively functionalized material depicts great potential for ultrasensitive and highly selective room temperature gas sensing, which is an ongoing challenge in chemiresistive and metal oxide gas sensors.

## 2. Materials and Methods

The device was fabricated using standard micro and nanofabrication techniques such as lithography, metal evaporation, and chemical vapor deposition CVD [8-11]. A 0.7mm thick Si substrate coated with 200nm SiO$_2$ is diced into 15mm*20mm substrates precleaned using acetone, IPA, and DI water. A 250 µm x 200 µm channel lithography was done for 5 tandem FETs on the substrate followed by 5nm Pt deposition using an E-beam evaporation. Direct tellurisation of the samples was performed using CVD. The deposited Pt films were placed in a two-zone CVD system. Samples were placed in 1$^{st}$ zone (300ºC) and tellurium powder in 2$^{nd}$ zone (430ºC). The image of placed glass slabs in furnace for CVD are shown in Fig 1. The tellurized samples were annealed in the furnace at 450ºC for 75 minutes with 100 sccm Ar gas.

Lithography and deposition were done for a 70µm×100µm area of the sensing layer.

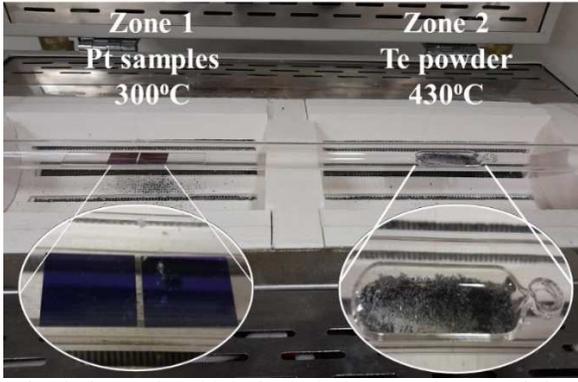

Fig 1. Chemical vapor deposition of Te in the furnace and two zoned placements of the substrate (Pt coated samples) at 300°C and material (Te powder) at 430°C

A composite thin film of $Au_{8nm}Pd_{6nm}$ was deposited using an E-beam evaporator as an ultra-selective sensing layer.

Finally, 100nm Au electrodes, adhering via 7nm Ti, were deposited using similar lithography and evaporation parameters. Finally, to test the fabricated sensors with target gas, a polydimethylsiloxane (PDMS) channel was fabricated and placed on the sensing layer of the tandem FETs to form a microfluidic channel.

### 3. RESULTS AND DISCUSSION

#### A. Electrical Characteristics

The electrical characteristics of the fabricated devices were tested using the Keithley and Agilent probe stations. The KickStart v2.0 software was employed to configure a voltage sweep spanning the gate and drain, while DinoCapture 2.0 facilitated real-time microscopic observation of the testing stage. Following the establishment of connections on the back gate, source, and drain, a voltage sweep was conducted ranging from -5V to +5V across the back gate ($V_{GS}$) and from -15V to +15V on the source-drain ($V_{DS}$). The resulting drain current ($I_D$) values were measured for each voltage, and a plot of the drain current against the drain voltage was generated to visualize the characteristic IV data points corresponding to each gate bias. Fig. 2 shows the forward and backward $V_{DS}$-$I_D$ curves for varying $V_{GS}$ of $PtTe_2$ FETs at room temperature (22°C). The gradual increase in the $I_D$ with $V_{DS}$ and $V_{GS}$ is symmetric and in line with classic FET-like IV characteristics.

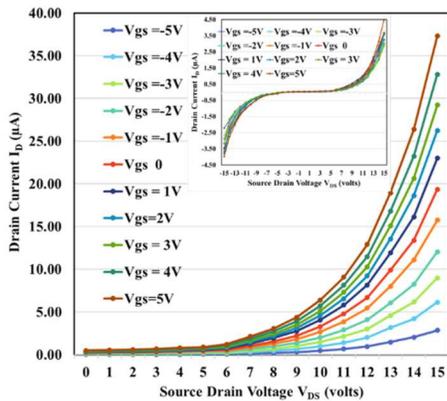

Fig 2. Stacked forward characteristic drain current ($I_D$) and drain-source voltage ($V_{DS}$) curves for $PtTe_2$ FETs and (inset) forward-backward $I_D$-$V_{DS}$

#### B. Structural Characteristics

The synthesized transducer material was structurally characterized using Scanning Electron Microscopy (SEM) and energy-dispersive X-ray spectroscopy (EDS). The SEM results, as shown in Fig 3 (a-c), depict the formation of $PtTe_2$ nanoparticles with distinct 2D structures observed at magnifications 150k, 300k and 600k. The well-defined structures at the nanoscale size and intricate patterns as exhibited in the SEM images showcase the crystalline nature of the synthesized material. The thickness of the material was also verified using atomic force microscopy and it was confirmed that 40nm thick $PtTe_2$ had been formed over the 5nm Pt after tellurisation.

The EDS chemometric analysis is listed in Table 1, which confirmed the presence of Pt and Te in significant ratios on the material growth region (tested regions of interest) as compared to the substrate. The EDX analysis reveals the elemental composition of Platinum ditelluride as 46.9% Tellurium (Te) and 32.3% Platinum (Pt), aligning with the stoichiometric ratio. The molar ratio of Te to Pt in Platinum ditelluride is approximately 1.45:1, confirming the expected fabrication of the 2D transducer compound ($PtTe_2$) based on elemental stoichiometry.

Table 1. Element stoichiometry of the synthesized channel material $PtTe_2$ using energy dispersive spectroscopy of the grown to confirm the elemental composition of Pt and Te in major ratios.

| Name | C | O | Si | Te | Pt |
|---|---|---|---|---|---|
| $PtTe_2$ Particles | 11.781 | 38.191 | 49.236 | 0.469 | 0.323 |
| Substrate | 6.567 | 42.902 | 50.531 | | |

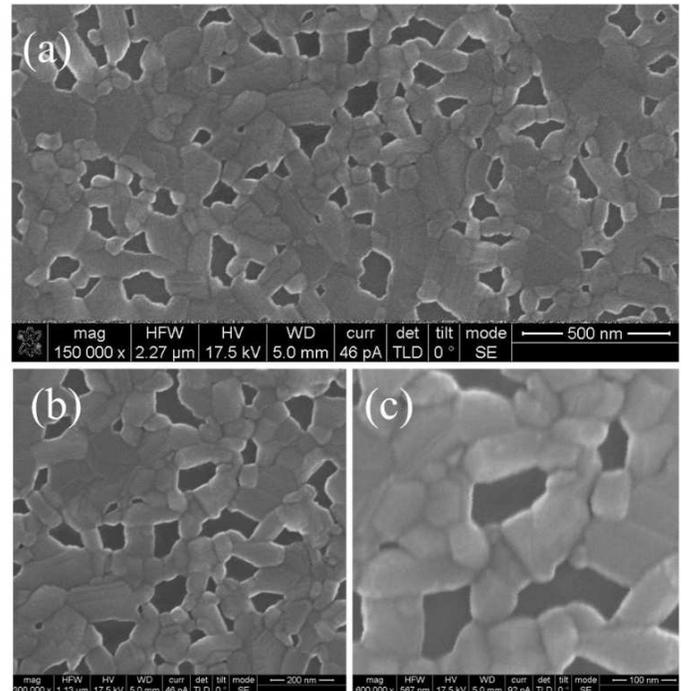

Fig 3. Scanning electron microscopy images of the 2D platinum ditelluride ($PtTe_2$) at magnifications (a) 150k, (b) 300k, and (c) 600k to demonstrate the nanoparticles.

## C. Ultrasensitive $H_2S$ Detection

The gas sensing setup consisted of gas cylinders ($N_2$ and $H_2S$), mass flow controllers (MFC), gas tubing with connectors, compressed air supply and data acquisition setup. The data acquisition setup consisted of Agilent Technologies source meter, a FET characterization probe station/board, QuickIV software on a laptop/PC wired with the source meter. The gas sensor is placed on the measuring station and 50ppm gas is flown through the PDMS channel on the sensor after acquiring a steady baseline. The gas is flown for 3 minutes and then the flow is ceased (so that the gas can stay uninterrupted on the surface) to acquire a steady state response of the sensor. Afterwards, compressed air is passed through the PDMS channel to flush out the $H_2S$. Fig 4 shows the transient response of the gas sensor in terms of $I_D$ at $V_G$= -4V; $V_{DS}$=1.05V. As evident from the sensor response, there is an 88.89% response of the sensor to the $H_2S$ at room temperature with a complete and unassisted recovery (also at R.T.), which is significant in the realm of gas sensors. It depicts the ultrasensitive nature of Au-Pd functionalized 2D $PtTe_2$ with longevity and stability at room temperature.

The Au and Pd composite thin film used for the sensing of $H_2S$ gas is highly selective and ultrasensitive for this particular analyte. The functionalization enables a high rate of physisorption of sulfide ions onto Pd with the assistance of Au and releases a higher number of electrons into the $PtTe_2$ channel upon exposure to the gas. This phenomenon leads to higher electric current (reduced resistance) in the presence of $H_2S$ gas. The sensing layer has significantly less or no affinity to other gases and hence the chemisorption or adsorption rate is very low, which makes it a selective gas sensor.

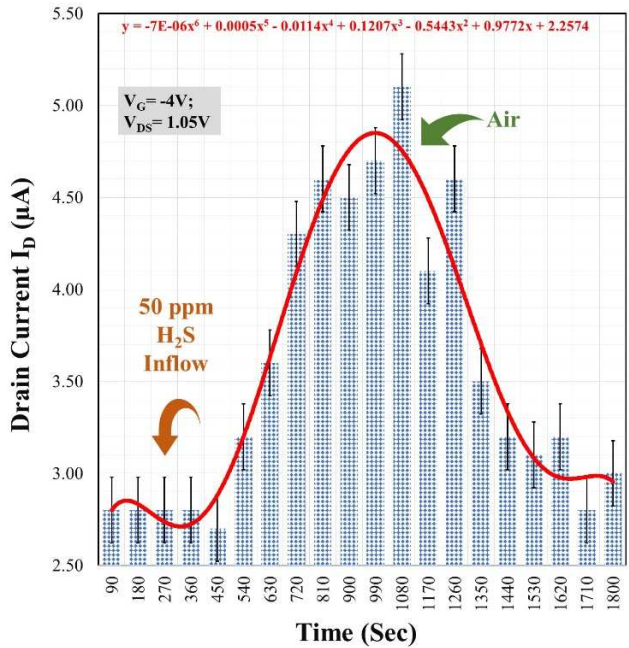

Fig 4. Drain current shift from baseline values to complete response upon 50ppm $H_2S$ exposure at $V_G$= -4V and $V_{DS}$=1.05V, followed by almost full recovery at compressed air ventilation.

## Conclusions

To epitomize, we demonstrated a unique FET-based gas sensor utilizing the emerging 2D $PtTe_2$ as a channel, decorated with noble metals for selective $H_2S$ detection. The material synthesis with its structural and electrical characteristics shows high mobility and makes it suitable for ultrasensitive chemical sensing. The Au-Pd composite thin film, used as a selective sensing material has depicted high and efficient response and recovery to $H_2S$ at room temperature and opened new avenues for further investigations.

## Acknowledgment

This work was performed in part at the Melbourne Centre for Nanofabrication (MCN) in the Victorian Node of the Australian National Fabrication Facility (ANFF).